\documentclass[prl,twocolumn,floatfix,amsfonts,showpacs]{revtex4}
\usepackage{graphicx,graphics,color,epsfig}
\usepackage{bm}
\usepackage{amsmath}
\usepackage{amssymb}
\begin{document}

\title{Spontaneous Emergence of Modularity in a Model of Evolving Individuals}
\author{Jun Sun and Michael W. Deem}
\affiliation{Department of Physics \& Astronomy and
Department of Bioengineering\\
Rice University, Houston, TX 77005--1892, USA}

\begin{abstract}
We investigate the selective forces that promote
the emergence of modularity in nature.
We demonstrate the spontaneous emergence of modularity in
a population of individuals that evolve in a
changing environment. We show that the level of modularity correlates with
the rapidity and severity of environmental change.
The modularity arises as a synergistic response to the noise in the
environment in the presence of horizontal gene transfer. 
We suggest that the hierarchical structure observed in the natural world may be
a broken symmetry state, which generically results from evolution
in a changing environment.  
\end{abstract}


\maketitle

Modularity abounds in biology.  Elements of hierarchy---modules---are
found in developmental biology, evolutionary biology, and 
ecology \cite{Shapiro2004, Shapiro2005, Lenski}.  Modularity 
is observed at levels that span
molecules, cells, tissues, organs, organisms, and societies.
At the genomic level, there are introns, exons, chromosomes, and genes.  
Moreover, there are mechanisms
to rearrange and transmit the information that is modularly
encoded at the genomic level, such as gene duplication, 
transposition, and
horizontal gene transfer
\cite{Shapiro,Goldenfeld2007}.
We define a module to be a component that can operate
relatively independently of the rest of the system. From a
structural perspective, existence of
modularity means there are more intra-module connections than
inter-module connections.  From a functional perspective, a
module is a unit that can perform largely the same function in
different contexts.
Modularity has been characterized in a variety of network
systems by physical methods \cite{Lipson2002, Wolde}. 
Selection for stability, for example, has been shown to select
for modular networks \cite{Lipson2004}.
A dictionary of constituent parts, or
network motifs, has been identified for 
the transcriptional regulation network
of \emph{E.\ coli} \cite{Alon2002}.
And once modularity has arisen, so that the goals a species
face become modular, modularly varying goals have been shown
to select for modular structure \cite{Alon2005}.
Horizontal gene transfer has been suggested to be essential 
to the evolution of a universal genetic code \cite{Goldenfeld2006}.

How does modularity arise \emph{a priori} in nature?
It has been suggested that by being modular,  a system
will tend to be both more robust to perturbations and more
evolvable \cite{Doyle, Kitano, Cluzel}.    
It has further been suggested that there is a selective
pressure for positive evolvability in a population of
individuals in a changing environment \cite{Earl}. 
Thus, we have hypothesized that modularity arises spontaneously
from the generic requirement that a population of individuals
in a changing environment be evolvable \cite{Deem2007}.
Support for this hypothesis has to date been elusive
\cite{Gardner2003}.

In this Letter, we show the hypothesis of spontaneous evolution
of hierarchy in a system under changing environmental conditions
to be valid.  Specifically, we show that in the presence of
horizontal gene transfer, environmental change leads to the
spontaneous emergence of modularity in a generic model of
a population of evolving individuals.  To represent the 
replication rate, or
microscopic fitness, of the individuals, we use a spin glass model
that has proved useful in previous studies of evolution
\cite{Kauffman,Deem,Sun}.   
To be specific, we choose parameter
values appropriate to describe a population
of evolving proteins \cite{Deem,Earl,Sun,Sun2006}.  
Spontaneous emergence of modularity, 
however, generically occurs for a population of evolving individuals and
depends only on the presence of a changing environment and the presence 
of horizontal gene transfer.
This spin glass model is appropriate because
it provides a rugged, difficult landscape upon which evolution struggles
to occur, and so there can be a pressure for more efficient
evolutionary structures to arise.  There are three time scales
in our system: the fastest time scale of sequence evolution
of the individuals in the population, the intermediate time scale
of environmental change, and the longest time scale of 
the change to the structure of protein fold space.
The symmetry of a uniformly random structure
is broken by the spontaneous emergence of modular structure as a response
to environmental change.

We use the following spin glass form
for the microscopic fitness of proteins
in our system (for a discussion on the spin glass
approach to evolution, see \cite{Deem,Earl,Sun,Sun2006}).
\begin{equation}
H^\alpha(s^{\alpha,k})=
\frac{1}{2\sqrt{N_D}}\sum_{i \neq j} \sigma_{i,j}(s^{\alpha,k}_i, s^{\alpha,k}_j) \cdot
\Delta_{i,j}^\alpha,
\label{1}
\end{equation}
where $s^{\alpha,k}_i$ is the amino acid
identity of the sequence $\alpha,k$ within fold $\alpha$ at position $i$,
and $N=120$ is the length of the protein sequence.   
We consider
the amino acids to lie within $5$ classes \cite{Deem}.
The term 
$\sigma_{ i,j}(s_i, s_j)$, is the interaction matrix,
symmetric in $i$ and $j$,
whose elements are each taken from
a Gaussian distribution with zero mean
and unit variance.  It differs for each
$i$, $j$, $s_i$, and $s_j$.  The effect of the environment
is encoded by these random couplings.  When the environment
changes with severity $p$, each of the couplings is
with probability $p$ randomly redrawn from the Gaussian distribution.
The term $\Delta_{ i,j}^\alpha$ defines the protein fold, i.e.\
the contact matrix, or connections in structure, for fold $\alpha$.
The matrix is
symmetric, with elements 0 or 1.
In order to guarantee that the emergence of modularity
comes from redistribution of connections rather than
an increase in the number of connections, 
we constrain $\sum_{ i > j+1} \Delta_{ i,j}^\alpha =
N_D=346$.  Any value of $N_D$ such that
the connection matrix is neither all unity nor all zero
would give qualitatively similar results.
We take $\Delta_{ i,i}^\alpha=0$ and $\Delta_{
i, i \pm 1}^\alpha=1$.

Because horizontal gene transfer will be assumed to transfer any of the
12 blocks of length 10 in the sequence, modularity is defined  
by the number of connections
within the 12 $10 \times 10$ blocks along the diagonal
\begin{equation}
M^\alpha=\sum_{k=0}^{11} \sum_{ i=1, j=i+2}^{10} \Delta_{10 k+i, 10k +j}^\alpha,
\label{eq:mod}
\end{equation}
so that $i,j$ are within the $1+k^{th}$ diagonal block of size
$10$.  Even a random distribution of contacts
will have a non-zero absolute modularity, $M_0$, and so
it is the excess modularity that measures the degree of
spontaneous symmetry breaking, $\delta M^\alpha=M^\alpha-M_0$.
Emergence of modularity means that as a result of evolution,
connections in structure are not evenly distributed
between positions.
The interactions are greater in the local, diagonal blocks
than in the rest of the matrix, and so
$\delta M^\alpha > 0$. 

In order to see the emergence of modularity, we need a set
of individuals in a changing environment.  Moreover, we need
a population of these sets, each set with a different
$\Delta_{i,j}^\alpha$.  We take the population size to be
$D_{\rm size}=300$ different structures, $1 \le \alpha \le D_{\rm size}$,
and each given structure has  a set of
 $N_{\rm size}=1000$
different sequences, $1 \le k \le N_{\rm size}$,
 associated with them. 
The average excess modularity is
given by $\delta M =  M - M_0 =
\frac{1}{D_{\rm size}}\sum_{\alpha=1}^{D_{\rm size}} M^\alpha - M_0$.

The structures, $\Delta_{i,j}^\alpha$,
are initialized by first randomly generating
one such structure with $N_D=346$ and a certain $M$. 
We then obtain the full set of $D_{\rm size}$ structures by
mutation away from this structure. 
Two elements of $\Delta_{ i,j}^\alpha$ with opposite
status are randomly chosen,
and the status of each is flipped from $1 \to 0, 0 \to 1$.
These mutations are done $n$ times, where $n$ is a Poisson
random number with mean 2.
The sequences,
 $s^{\alpha, k}_i, 1 \le i \le N$, of each individual
are initialized by random assignment.

The
evolution in our simulation involves three levels of change.
The most rapid change occurs by evolution of the sequences
through point mutation and gene segment swapping.
For
each structure $\Delta_{ i,j}^\alpha$, at each round, all the $N_{\rm
size}$ associated sequences undergo point mutation,
gene segment swapping,
and selection.  The Poisson point mutation process 
changes on average $2.4$ amino acids per sequence, which are randomly
selected and assigned a random class. 
In gene segment swapping two
randomly selected sequences from the population associated with one
structure attempt to exchange each of the 12
sequence fragments between $10k+1$ and $10k+10$
(of length $10$) with probability $0.1$. 
The qualitative behavior of the results does not depend on the exact
mutation rates.
Pairs of sequences in the population associated with one structure
are chosen, until all sequences have been chosen.
This process is a model 
of horizontal gene transfer and recombination. 
The 50\%\ sequences with the lowest energy 
are selected and randomly duplicated to recover the population of
$N_{\rm size}$ for the next round; the
microscopic replication rate, or fitness, for sequence $\alpha,k$ in
structure $\alpha$ is
$r^\alpha(s^{\alpha,k}) = 
2 \theta [H^\alpha_{{\rm N_{size}}/2} - H^\alpha(s^{\alpha,k})]$, where
$\theta(x)$ is the Heavyside step function.
Mutation and
selection are repeated $T_2$ rounds.

The next most rapid change is that of the environment,
which occurs with severity $p$ and frequency $1/T_2$.
During the environmental change, the 
elements of the interaction matrix
$\sigma_{i,j}$ change with probability $p$.

The slowest level of change is the structural evolution.
The selection at this level is based on
the cumulative fitness of the set of individuals with a given structure,
averaged over $T_3 = 10^4 T_2$ environmental changes. 
The structures with the best $5\%$
cumulative fitness are selected and randomly amplified to
make the new population of $D_{\rm size}$ structures, $\Delta^\alpha_{ij}$.
The structure
population also undergoes mutation. 
As with the initial construction,
two elements of $\Delta_{ i,j}^\alpha$ with opposite
status are randomly chosen,
and the status of each is flipped from $1 \to 0, 0 \to 1$.
These mutations are done $n$ times, where $n$ is a Poisson
random number with mean 2.
The mutated
structures, $\Delta_{ i,j}^\alpha$,  are used for the next
$T_3$ rounds of evolution.

\begin{figure}[t]
\begin{center}
\epsfig{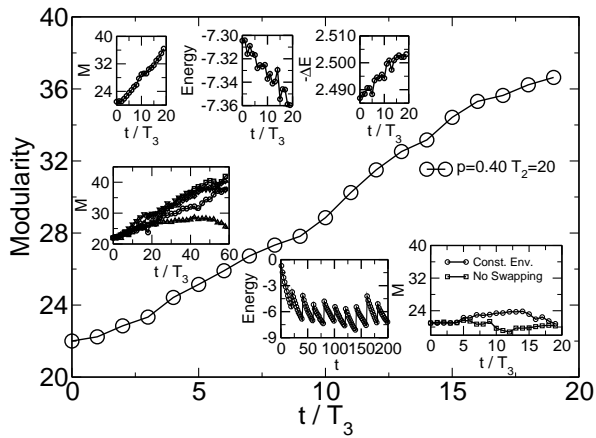}
\end{center}
\caption{Spontaneous emergence of excess modularity,
$M > M_0 = 22$ from
a state with no excess modularity, 
$M=M_0$. The random, symmetric distribution of
structural connections is spontaneously broken as system
evolves. Here $T_2=20$, and the severity of environmental change is 
$p=0.40$. 
The upper left inset shows the growth of modularity starting
from a power law distributed contact matrix ($\gamma=3$).  The upper
middle inset shows the improvement in the energy as the modularity
grows.  The upper right inset shows the improvement of evolvability, 
or change in energy in one environment, as the modularity grows.
The inset in the middle row shows the emergence of modularity as a 
result of a horizontal gene transfer operator with a Poisson random 
swap length 
and uniform random starting position. 
Shown are data for an average swap length of 
10 ($\bigcirc$),
20 ($\Box$), 
20 ($\Diamond$), 
5 ($\bigtriangleup$),
and 40 ($\bigtriangledown$)
 with 12, 6, 12, 24, and 3 attempted swaps, respectively,
of probability 0.1 per sequence pair.
The lower left inset shows how the energy changes within
one environment and between environmental changes ($T_2=20$).  
The lower right
inset shows that emergence of modularity requires both environmental
change and horizontal gene transfer. 
In all cases modularity is measured by Eq.\ \ref{eq:mod}, and excess modularity
is measured by $\delta M^\alpha=M^\alpha-M_0$, with $M_0=22$. 
}
\label{fig:MandT-O}
\end{figure}

In Fig.~\ref{fig:MandT-O} we show the spontaneous emergence of
modularity from the symmetric, random state 
of no excess modularity, $M = M_0 = 22$.
Since the system is initially quite far from the steady state modularity,
the growth of the excess modularity with time is roughly linear.
The excess modularity is the order parameter for this system, and its 
growth shows that the system is in a broken symmetry phase
with modular structure under these conditions.
Interestingly, the growth of modularity is identical for an initial
contact matrix that is power-law distributed with $\gamma=3$.  
Figure \ref{fig:MandT-O}  shows that emergence of modularity in this model
requires both horizontal gene transfer and a changing environment.  
The spontaneous emergence of modularity is a general result. 
In Fig.~\ref{fig:MandT-O}, we show the excess modularity still grows, even if 
the gene transfer starts at a uniformly random position and
swaps a random length of sequence.
The original assumption of fixed length and position,
however, is biologically motivated.  If the blocks are exons, and
the ratio of non-coding to coding DNA is large, then typical recombination
or horizontal gene transfer will transfer an integer number of complete
blocks, which is our horizontal gene transfer operator
of fixed length and position. 

\begin{figure}[t]
\begin{center}
\epsfig{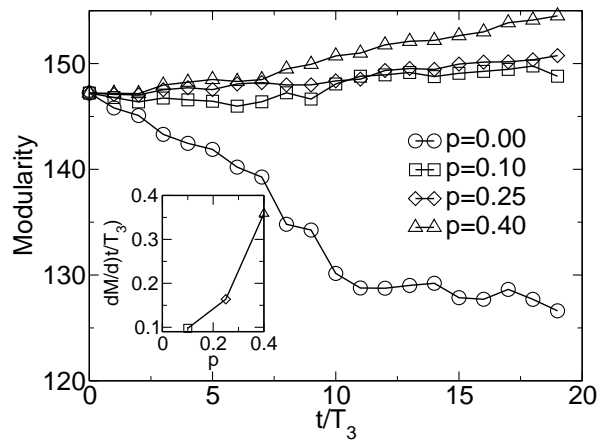}
\end{center}
\caption[]{The velocity at which modularity grows is positively
correlated with the magnitude of environment change, $p$. The
frequency of environment change is set at $1/T_2=1/40$.  The inset
shows the response function of the system $dM/d (t/T_3)$ as a function of
the severity of environmental change.}
\label{fig:MandT-A}
\end{figure}
The system adopts the broken-symmetry, modular state not because
the point mutation and gene segment swapping moves favor modularity
a priori, but rather because these moves enable the system to respond
more effectively to a changing environment when the system is modular.
That is, evolvability is implicitly selected for in a changing environment, and
gene segment swapping enhances evolvability if the system is modular.
Thus, we expect modularity to be implicitly selected for in a changing
environment in the presence of horizontal gene transfer,
with the degree of modularity correlated to the degree of environmental
change.
In Fig.~\ref{fig:MandT-A} we show the change of modularity with time
for different severities of environmental change, $p$.
For this figure, we choose the initial set of structures from
an ensemble with
$M=147$, rather than $M=M_0$, to show the change of
modularity more clearly.
For no environmental change, the modularity decreases
from this high level.  But for positive environmental change, the
modularity increases from the initial, high level.  The velocity of the
increase is larger for greater environmental change.

\begin{figure}[t]
\begin{center}
\epsfig{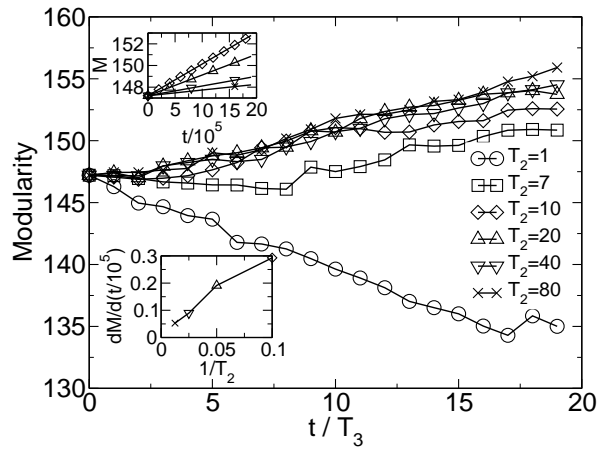}
\end{center}
\caption{Frequency of environmental change also affects the
time evolution of spontaneous modularity.
The severity of environment change is $p=0.40$.
The upper inset shows the growth of modularity versus real time rather
than versus time relative to the number of environmental changes.
The lower inset  shows
the response function of the system $dM/d (t/10^5)$ as a function of
the frequency of environmental change.}
\label{fig:MandT-F}
\end{figure}

Another way of characterizing the environmental change is by the
frequency of change, and the emergence of modularity depends on this
parameter as well.
In Fig.~\ref{fig:MandT-F} we show the growth of modularity with
time for different frequencies of environmental change.
For frequencies of environmental change that are not too large, the
modularity increases with frequency.  For very high frequencies,
$1/T_2 > 1/5$, the system is unable to track the changes in the environment,
and the modularity decays with frequency.
The velocity of modularity increase in 
Fig.~\ref{fig:MandT-F} 
for $p=0.40$ and $T_2=20$ is less than that in
Fig.~\ref{fig:MandT-O} because in 
Fig.~\ref{fig:MandT-F},  the system is
closer to the steady-state, broken-symmetry value than it is in the
Fig.~\ref{fig:MandT-O}.

The spontaneous emergence of modularity is caused by the historical
variation in environments that the system has encountered.  By
a fluctuation-dissipation argument \cite{Hanngi1998,Kaneko,Earl},
 we might expect that
the degree of modularity should be proportional to the variance of
environments encountered.
In the inset to Fig.\ \ref{fig:MandT-A} we show that the
velocity of the increase in modularity is roughly proportional to the
severity of environmental change, $p$.
In the inset to Fig.\ \ref{fig:MandT-F} we show that the
velocity of the increase in modularity is roughly proportional to the
frequency of environmental change, $1/T_2$.

\begin{figure}[t]
\begin{center}
\epsfig{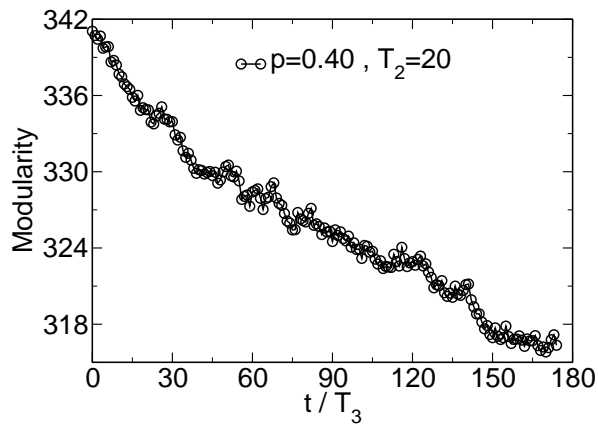}
\end{center}
\caption[]{
If the initial value of the modularity is greater than the
steady-state value, the modularity decays with time.
Here $T_2=20$, and the severity of environment change is 
$p=0.40$.  
}
 \label{fig:MandT-S}
\end{figure}
While the modularity grows with time in
Figs.\ \ref{fig:MandT-O}--\ref{fig:MandT-F} for
$p>0$ and $T_2 >5$, at steady state the system will
be only partially modular, $M < N_D=346$, 
reflecting a balance between the selection
for modularity in a changing environment and 
the mutations driving the system toward the symmetric state
of no excess modularity.
See Fig.~\ref{fig:MandT-S}.
The excess modularity in the broken symmetry state is
positive because of
selection for modularity in fluctuating environments, and the
excess modularity is not the maximal possible value 
of $M=N_D=346$
because of the entropic effects of the mutations in the sequence space.
For the initial condition used in Fig.~\ref{fig:MandT-S},
nearly all the connections in the diagonal blocks
and few in the off-diagonal blocks,
modularity decays over time, showing the steady state value is
below $316$. The modularity will saturate at a value for which
the effects of selection pressure and 
mutation balance each other.
Further experimental study of the relation between
large scale genetic exchange and the promotion of modularity
is warranted \cite{Lenski}. 
Some species of yeast may undergo either sexual or asexual
reproduction, and experiments suggest that yeasts undergoing
sexual reproduction are more evolvable \cite{Burt}. It would
be interesting to construct protocols to study the relation between sexual
recombination and modularity, possibly in gene expression
networks \cite{Bonhoeffer}, in the laboratory.
At an applied level, we note that
the process by which antibiotics resistance evolved \cite{Walsh}
makes use of the modular structure of the genes encoding
the enzymes that degrade and the pumps that excrete antibiotics
and the modular structure of the proteins to which
antibiotics bind \cite{Maiden}.

Why is modularity so prevalent in the natural world?  Our hypothesis is
that a changing environment selects for
adaptable frameworks, and competition among different evolutionary frameworks
leads to selection of structures with the most efficient dynamics,
which are the modular ones.  
We have provided evidence validating this hypothesis.
We suggest that the beautiful, intricate, and interrelated structures
observed in nature may be the generic result of 
evolution in a changing environment.
The existence of such structure need not necessarily
rest on intelligent design or the anthropic principle.

\vspace{-0.25in}

\bibliography{modularity}

\end{document}